%% file: main.tex
\documentclass[journal]{IEEEtran}
\IEEEoverridecommandlockouts
\usepackage{cite}
\usepackage{amsmath,amssymb,amsfonts,upgreek}
\usepackage{algorithmic}
\usepackage{textcomp}
\usepackage{xcolor}
\usepackage{cite}
\usepackage{url}
\usepackage{amsfonts}
\usepackage{amssymb}
\usepackage{mathrsfs}
\usepackage{euscript}
\usepackage{graphicx}
\usepackage{multirow}
\usepackage{algorithm}
\usepackage{algorithmic} 
\usepackage{changepage}
\usepackage{color} 
\usepackage{caption}
\usepackage{subcaption}
\usepackage{todonotes}
\usepackage{scalerel}
\usepackage{tikz} 
\usepackage{pgfplots} \usepgfplotslibrary{groupplots} \pgfplotsset{compat=1.18}
\usetikzlibrary{svg.path}
\usepackage{balance}
\usepackage{svg}
\usetikzlibrary{fit, positioning, shapes.geometric, fit, arrows.meta, calc, backgrounds}
\usepackage{pgfplots}
\pgfplotsset{compat=newest}
\definecolor{darkgreen}{rgb}{0, 0.5, 0} 
\definecolor{lightpurple}{rgb}{0.7, 0.4, 1} 
\definecolor{orcidlogocol}{HTML}{A6CE39}
\tikzset{
orcidlogo/.pic={
\fill[orcidlogocol] svg{M256,128c0,70.7-57.3,128-128,128C57.3,256,0,198.7,0,128C0,57.3,57.3,0,128,0C198.7,0,256,57.3,256,128z};
\fill[white] svg{M86.3,186.2H70.9V79.1h15.4v48.4V186.2z}
svg{M108.9,79.1h41.6c39.6,0,57,28.3,57,53.6c0,27.5-21.5,53.6-56.8,53.6h-41.8V79.1z M124.3,172.4h24.5c34.9,0,42.9-26.5,42.9-39.7c0-21.5-13.7-39.7-43.7-39.7h-23.7V172.4z}
svg{M88.7,56.8c0,5.5-4.5,10.1-10.1,10.1c-5.6,0-10.1-4.6-10.1-10.1c0-5.6,4.5-10.1,10.1-10.1C84.2,46.7,88.7,51.3,88.7,56.8z};
}
}
\newcommand\orcidicon[1]{\href{https://orcid.org/#1}{\mbox{\scalerel*{
\begin{tikzpicture}[yscale=-1,transform shape]
\pic{orcidlogo};
\end{tikzpicture}
}{|}}}}
\usepackage[colorlinks=true,linkcolor=blue,citecolor=blue]{hyperref}

\def\BibTeX{{\rm B\kern-.05em{\sc i\kern-.025em b}\kern-.08em
T\kern-.1667em\lower.7ex\hbox{E}\kern-.125emX}}
\usepackage[most]{tcolorbox}

\begin{document}

\author{
$ 
\text{Jalal Jalali}^{\orcidicon{0000-0002-3609-6775}}\ \IEEEmembership{Member, IEEE},$
$\text{Mostafa Darabi}^{\orcidicon{0000-0003-4194-3307}} \  \IEEEmembership{Member, IEEE},$
\text{and}\ 
$ \text{Rodrigo C. de Lamare}^{\orcidicon{0000-0003-2322-6451}}\ \IEEEmembership{Fellow, IEEE}.$ \vspace{-1.5em}
\thanks{Jalal Jalali and Rodrigo C. de Lamare are affiliated with the Centre for Telecommunications Studies, Department of Electrical Engineering (DEE), Pontifical Catholic University of Rio de Janeiro, Rio de Janeiro 22451900, Brazil (e-mail: delamare@puc-rio.br).
Jalal Jalali and Mostafa Darabi are with the Wireless Communication Research Group, JuliaSpace LLC., Chicago, IL, USA. All authors contributed equally to this work.
}
\thanks{This work was supported by the São Paulo Research Foundation, FAPESP, under Grant 2024/14280-0. \textcolor{black}{The source code is publicly available on GitHub:} \textcolor{black}{\href{https://github.com/mostafadarabi/Shape-Adaptive-RHS}{https://github.com/mostafadarabi/Shape-Adaptive-RHS}}.}}

\title{\huge Shape Adaptive Reconfigurable Holographic Surfaces}

\maketitle
\begin{abstract}
The evolution of wireless networks toward sixth-generation (6G) is driving demand for intelligent, flexible, and energy-efficient propagation control. Reconfigurable holographic surfaces (RHSs) enable fine-grained wavefront manipulation through densely packed programmable elements. \textcolor{black}{This paper introduces shape-adaptive RHSs, which dynamically change surface geometry to provide extra spatial degrees of freedom. The adaptive design responds to environmental dynamics, mitigates blockages, and improves performance across urban communications, aerial coverage, and joint sensing and communication scenarios.} We propose a network architecture that jointly designs the surface shape at the RHS and optimizes transmission. Simulation results show that the proposed shape-adaptive RHSs significantly increase data rates. These findings establish shape-adaptive RHS as a foundational technology for intelligent, context-aware wireless environments in 6G and beyond wireless networks.

\end{abstract}
\section{Introduction}
\IEEEPARstart{D}{elivering} high data rates and reliable connectivity at millimetre-wave (mmWave) and terahertz (THz) frequencies remains one of the central challenges for sixth-generation (6G) wireless networks because of severe path loss, blockage sensitivity, and limited diffraction~\cite{9410457}. Although large directional antenna arrays can compensate for these propagation impairments, they require a large number of radio-frequency (RF) chains, phase shifters, and power amplifiers, resulting in increased hardware complexity, cost, and energy consumption~\cite{10530348}. These limitations have motivated the development of programmable metasurfaces that actively control the wireless propagation environment rather than relying solely on increasingly complex transceiver architectures.

Among these technologies, reconfigurable intelligent surfaces (RISs) have attracted considerable attention~\cite{10737121}. By employing large arrays of low-cost programmable elements, RISs manipulate the phase of externally incident electromagnetic (EM) waves without requiring a dedicated RF chain for each element~\cite{10737121}. 
This enables energy-efficient beam steering and has stimulated extensive research on coverage enhancement, interference management, physical-layer security, integrated sensing and communication, and energy-efficient wireless networks. 
Recent studies have further investigated RIS-assisted systems through reflection-coefficient optimization~\cite{10606173}, multi-RIS deployment and selection~\cite{10139787}, coverage extension~\cite{345678765433} partial element activation for energy savings~\cite{10373958}, unmanned aerial vehicle (UAV)-assisted communications~\cite{10129204}, and secure edge computing and content delivery~\cite{9416239}.

Despite these advances~\cite{10737121,10606173,10139787,345678765433,9416239,10129204,10373958}, existing RIS architectures share a common characteristic: the physical aperture remains fixed. \textcolor{black}{Whether passive, active, beyond-diagonal, or multilayer/stacked, RIS designs primarily optimize the EM response of a predefined planar surface by adjusting reflection coefficients~\cite{10606173}. Consequently, the geometry of the active aperture itself is not exploited as a design variable.} At higher frequencies, however, wireless propagation is highly sensitive to blockage, user location, and spatial geometry, suggesting that adapting the effective aperture can provide additional spatial degrees of freedom beyond conventional phase and amplitude control.

\textcolor{black}{This observation naturally motivates the evolution toward reconfigurable holographic surfaces (RHSs)~\cite{rhs_holographic_radio_2023}. Unlike RISs, which passively reflect externally incident waves, RHSs integrate one or more embedded feeds with densely arranged programmable leaky-wave elements that directly synthesize EM wavefronts according to the holographic principle~\cite{9696209}. 
Because beamforming is achieved through the controlled radiation of surface waves rather than through a dedicated RF chain for every element, RHSs provide a highly efficient large-aperture transceiver architecture while offering much finer control over the radiating aperture~\cite{10163760}.}

\textcolor{black}{Building upon these capabilities, this paper introduces the concept of shape-adaptive RHSs. Our proposed approach changes the geometry of the effective radiating aperture by activating different subsets of meta-elements. This additional spatial degree of freedom enables the propagation environment to be adapted to adverse network conditions while preserving the simplicity of a planar implementation.} This work aims to pave the way for deploying intelligent metasurfaces that are not only reconfigurable in the signal domain but also geometrically adaptive, offering extra degrees of freedom for sustainable high-performance systems in the 6G era.

\section{Fundamentals of Shape-Adaptive RHSs}
\subsection{From RISs to RHSs: Core Differences}
{\color{black}{Although RISs and RHSs are both programmable metasurfaces, they manipulate EM waves through fundamentally different mechanisms. A conventional RIS behaves as a programmable reflector: an incident EM wave illuminates the surface, and each sub-wavelength meta-element adjusts its reflection coefficient to modify the phase, and in some implementations, the amplitude, of the reflected field~\cite{10737121}. Consequently, the transmitted energy originates entirely from an external source, and the RIS can only redistribute the impinging wave over its fixed physical aperture.

An RHS follows a different operating principle. Rather than relying on external illumination, it integrates one or more feeding structures beneath the metasurface to launch guided reference waves that propagate across the surface~\cite{9696209}. As these waves travel, programmable radiation elements gradually leak energy into free space, where the radiated fields combine constructively in desired directions and destructively elsewhere to synthesize directive beams according to holographic and leaky-wave radiation principles~\cite{rhs_holographic_radio_2023}. Since beamforming is realized by controlling how the guided wave radiates across the surface, rather than by independently driving every element with dedicated RF chains, RHSs provide an efficient approximation of a continuous radiating aperture. More importantly, this radiation mechanism naturally enables selective activation of the metasurface, allowing only specific regions to participate in beam formation while the remaining elements remain inactive. 
Consequently, the effective radiating aperture itself becomes programmable, introducing an additional degree of freedom beyond conventional beamforming: \textit{shape adaptation}. This capability is realized by treating the spatial support of the radiating aperture as an optimization variable. Instead of activating every meta-element, the controller selects a two-dimensional subset of elements from a predefined shape codebook. Each shape, therefore, defines a different effective aperture while preserving the same feeding network and hardware architecture. The conventional full-aperture RHS is recovered as a special case when all elements are activated~\cite{10163760}. 

Shape adaptation in RHSs does not involve mechanically bending the panel, introducing surface curvature, or modifying the printed circuit board (PCB). The physical RHS hardware may remain completely planar throughout operation. Shape adaptation is achieved electronically by changing which subset of meta-elements participates in radiation~\cite{rhs_holographic_radio_2023}. 
As a result, existing planar manufacturing techniques, installation procedures, and sectorized deployments stay unchanged, while the EM aperture can still adapt dynamically to different propagation conditions. From a system perspective, this extra degree of freedom complements conventional beamforming. Besides controlling the radiated EM field, the network can determine where radiation is generated across the surface itself. Consequently, the aperture geometry becomes another optimization variable that can be jointly designed with beamforming and resource allocation to better accommodate network objectives.}}

\subsection{Structure and Working Principle of RHSs}
An RHS is typically organized into three tightly integrated functional layers. The bottom layer contains the feeding network, which may consist of substrate-integrated waveguides (SIWs), microstrip transmission lines, or other guided-wave structures that inject EM energy into the metasurface. Above the feeding network lies a dense array of programmable meta-elements incorporating tunable components, e.g., varactor diodes, PIN diodes, or micro-electromechanical systems (MEMS), that locally control the EM response. Together, these layers produce the desired far-field pattern~\cite{10080950}.
Holographic radiation is obtained by programming the surface response so that the guided reference wave and radiated fields combine constructively in desired directions. This control approximates a continuous aperture and enables directive, smooth beam patterns without requiring a dedicated RF chain and phase shifter for every element.

Shape adaptation selectively activates geometric subsets of the surface. A controller maps channel conditions, user locations, and network objectives to an activation mask and radiation coefficients, jointly controlling the emitted wavefront and effective aperture. This closed-loop operation allows the active region to follow changes in the propagation environment without modifying the physical panel. Inactive regions remain non-radiating and may also support other functionalities, e.g.,  sensing, energy harvesting, or interference suppression.

\begin{figure*}
    \centering
    \vspace{-5mm}
    \includegraphics[width=0.95\linewidth]{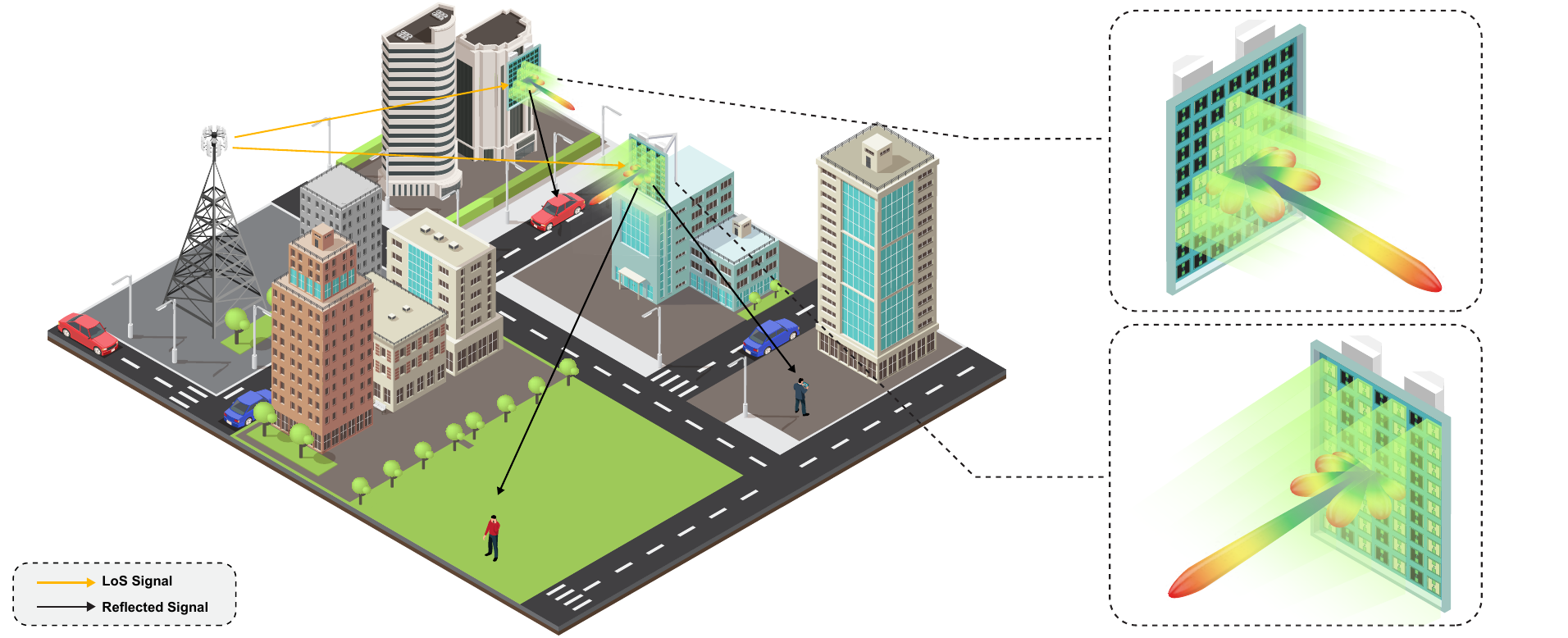}
    \caption{Shape-adaptive multi-RHS-assisted network architecture. The figure illustrates a downlink communication system in which a BS serves multiple users via multiple RHS. Each RHS consists of a planar array of elements, with different activation patterns optimized for network performance. The left RHS follows a rectangular activation shape, while the right RHS utilizes a column-wise striped activation shape. These shape variations influence the {\color{black}holographic beamforming} and impact signal propagation.}
    \vspace{-3mm}
    \label{fig:sysmodel}
\end{figure*}

\subsection{Shape Adaptation: A New Degree of Freedom} 
In conventional RISs, beamforming relies on adjusting phase shifts across a static flat grid of elements, akin to configuring a fixed passive beamformer~\cite{10737121}. 
Unlike continuous amplitude tapering, which independently adjusts the radiation amplitude of each element, discrete shape-codebook switching selects among predefined spatial activation patterns, thereby changing the geometric support of the effective radiating aperture. 
Accordingly, shape-adaptive RHSs introduce extra degrees of freedom by allowing selective activation of geometric sub-regions, changing the surface shape to alter radiation characteristics. This mechanism can be interpreted similarly to codebook-based beamforming in multiple-input multiple-output (MIMO) systems, where each pre-defined codeword corresponds to a beam pattern optimized for a specific user direction or channel condition~\cite{rhs_holographic_radio_2023}. 

\textcolor{black}{In shape-adaptive RHS, each predefined shape configuration (e.g., rectangular, L-shaped, U-shaped, or diagonal patterns, cf.~Fig.~\ref{fig:activation_patterns}) forms a discrete member of a spatial codebook.} These configurations control not only the active element locations but also their geometric layout, which in turn modifies the angular spectrum and beamwidth of the radiated signal. 
\textcolor{black}{Similar to codebook-based systems that scan different beam patterns to improve signal strength or reduce interference~\cite{10071555}, the RHS controller can select from a set of predefined shape configurations, that is, a dictionary, to optimize a given network performance metric, e.g., maximize the signal-to-noise ratio (SNR).} 
This shape-codebook approach reduces the computational burden of fully continuous optimization while preserving flexibility to dynamically adapt to environmental changes. Moreover, this enables rapid adaptation through low-latency switching among shape patterns. Unlike traditional RISs, which only steer incoming signals, RHSs with shape adaptation actively co-design the spatial support of the outgoing beam, enabling finer-grained, more responsive control over the propagation environment.

\subsection{Practical Constraints for Shape-Adaptive RHSs}
Shape adaptation in RHSs must satisfy power leakage constraints so that EM energy is radiated toward intended users with limited unintended dispersion~\cite{rhs_holographic_radio_2023}. Since RHSs rely on surface-wave excitation and leaky-wave radiation, poorly configured shapes may leak power too early, reduce beamforming gain, or cause interference, while insufficient leakage traps energy and lowers radiation efficiency. Therefore, each meta-element should keep its local impedance, excitation amplitude, or radiation coefficient within predefined bounds to balance confinement and radiation.

Element-level control enforces this balance through tunable components, e.g., MEMS switches, which regulate each element's contribution to surface-wave perturbation and radiation~\cite{10080950}. By adjusting these parameters, the RHS controller sets how much power each element extracts and radiates while satisfying system and regulatory limits. In shape-adaptive RHSs, where only a subset of the surface is active at a given time, the power distribution must be rebalanced after each shape change to avoid localized over-radiation or under-utilization. This can be achieved by normalizing the feed excitation or adapting the radiation coefficients so that the total leaked power remains consistent across shape configurations.

Accordingly, power leakage constraints can be modelled as hard limits for EM compliance and stable operation, or as soft penalties that trade efficiency, beam sharpness, and coverage flexibility. Overall, power-aware surface programming with precise element-level control enables dynamic geometric reconfiguration while maintaining efficient, stable, and interference-aware operation in dense wireless deployments.

\section{Prototype and Implementation Considerations}
While shape-adaptive RHSs offer significant theoretical and system-level advantages, their practical realization depends on advances in metasurface fabrication, control electronics, and system integration. This section discusses (\ref{III_a}) the feasibility of implementing dynamic element control, (\ref{III_b}) the associated energy efficiency and cost benefits, and (\ref{III_c}) the potential for integration with existing wireless infrastructure.

\subsection{Feasibility of Dynamic Element Control}\label{III_a}
Recent progress in programmable metasurfaces has demonstrated the practical feasibility of dynamically controlling large numbers of meta-elements in real time~\cite{rhs_holographic_radio_2023,10163760,9696209}. 
RHS implementations typically rely on tunable components to adjust the local EM response of each element. These technologies have already been validated in RIS and holographic metasurface prototypes operating from sub-6 GHz up to mmWave and THz frequencies~\cite{9410457}. 
For shape-adaptive RHS, dynamic element control extends beyond phase or amplitude tuning to include selective activation and deactivation of spatial regions across the surface. 
\textcolor{black}{This can be realized through low-power switching networks that enable groups of elements to be reconfigured as contiguous geometric shapes. The considered implementation relies on electronic activation and deactivation of meta-elements within a fixed planar panel and does not require mechanical deformation of the surface.}


The reconfiguration timescale of metasurface controllers, ranging from microseconds to milliseconds, is well matched to the coherence times of practical wireless channels. 
This enables the activation mask to be updated in response to channel conditions without incurring excessive latency. 
As a result, dynamic shape adaptation can be practically implemented as part of closed-loop wireless control systems.

\subsection{Energy Efficiency and Hardware Cost Benefits}\label{III_b}
The energy efficiency of shape-adaptive RHSs stems primarily from activating only the portion of the aperture that is required for a given communication task. Instead of uniformly energizing the entire surface, the controller can allocate the available radiated power to the most effective aperture configuration, reducing unnecessary energy expenditure while maintaining the desired coverage or beamforming objective. The inactive regions consume only their biasing and control power and may simultaneously support auxiliary functionalities.

\textcolor{black}{From an implementation perspective, holographic beamforming introduces a different hardware trade-off from conventional antenna arrays. To represent the phase progression of the guided reference wave with an approximate phase resolution of $\pi/2$, the spacing between adjacent meta-elements along the propagation direction, $d$, is generally chosen to satisfy $d\leq\lambda_{\mathrm{g}}/4$, where $\lambda_{\mathrm{g}}$ denotes the guided wavelength. Consequently, a given aperture generally contains more programmable elements, biasing lines, and control connections than RIS or phased array, while a denser arrangement may increase mutual coupling and calibration requirements.}

\textcolor{black}{Nevertheless, these additional elements are implemented using relatively simple tuning devices rather than complete RF front ends. Therefore, the implementation complexity of an RHS is governed by a trade-off between a higher element density and a substantially lower per-element hardware complexity, instead of by the number of RF chains alone~\cite{rhs_holographic_radio_2023}.} This trade-off makes large-aperture RHS deployments an attractive option where conventional massive MIMO arrays would become prohibitively expensive or power-hungry.

\subsection{Potential Integration with Existing Infrastructure}\label{III_c}
Shape-adaptive RHSs can be seamlessly integrated into existing wireless infrastructure thanks to their thin, lightweight, and conformal structure. They can be installed on building facades (cf.~Fig.~\ref{fig:sysmodel}), indoor surfaces, lamp posts, or transportation infrastructure without major structural changes, enabling cost-effective coverage enhancement in dense urban areas. Beyond static deployments, RHS panels can also be mounted on UAVs, vehicles, or temporary platforms to provide on-demand connectivity, where their low power consumption and compact design are particularly advantageous.

\textcolor{black}{Network coordination can be realized through existing control and backhaul links, allowing shape-adaptive RHS controllers to receive configuration updates from the serving base station (BS) or a centralized network controller. Consequently, the shape-adaptive RHSs can be introduced gradually alongside existing cellular and Wi-Fi infrastructure, providing additional propagation flexibility while remaining backward compatible with current network deployments.}

\section{Shape Selection and Radiation Characteristics}
\begin{figure*}[t]
    \centering
    \resizebox{0.80\width}{!}{%
    \input{rhs_shapes.tex}
    }
\caption{Representative activation patterns for a $16\times16$ shape-adaptive RHS. Black cells denote active (ON) elements, while white cells denote inactive (OFF) elements. Different activation geometries produce distinct effective apertures and radiation characteristics.}
    \label{fig:activation_patterns}
\end{figure*}
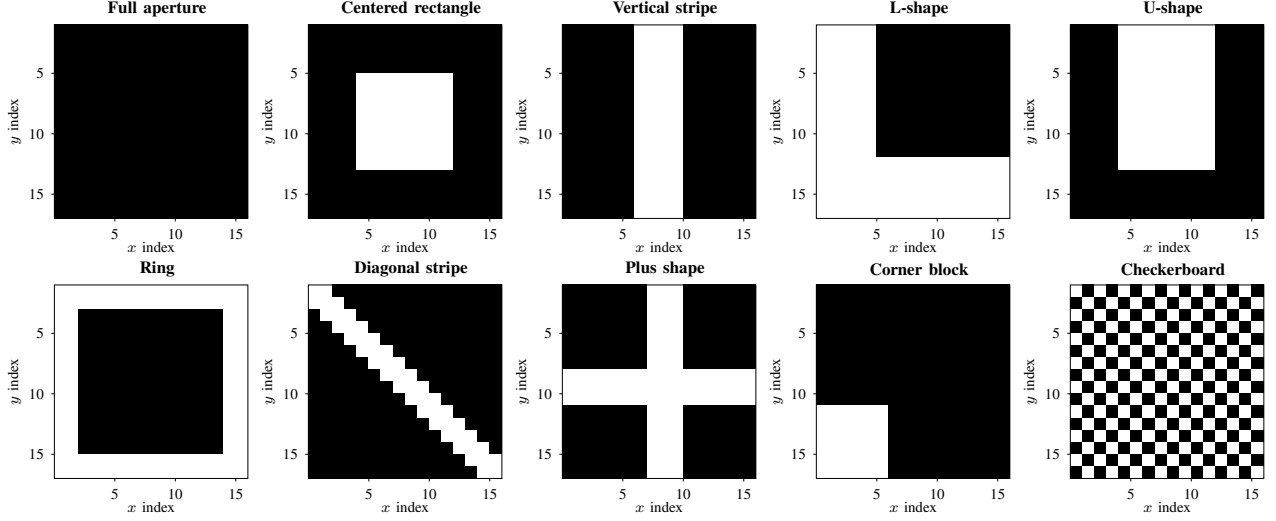

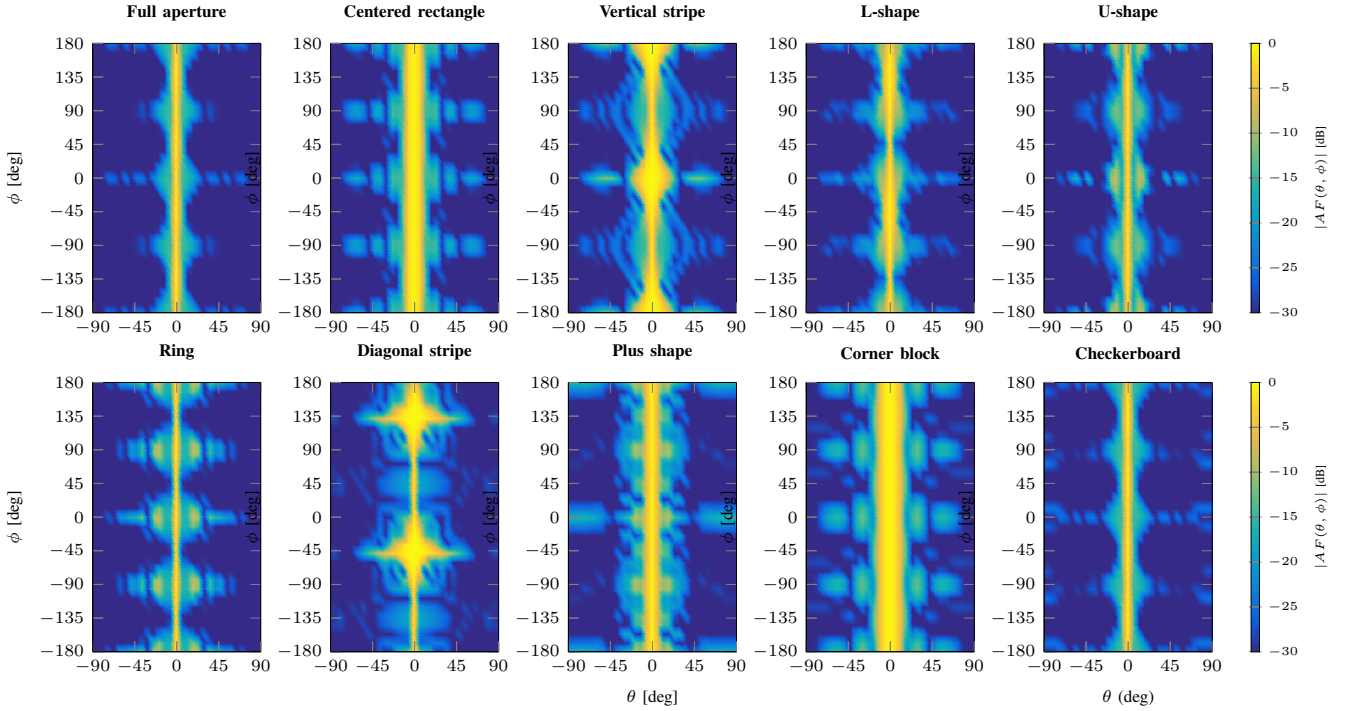
\begin{figure*}[t]
    \centering
    \resizebox{\textwidth}{!}{
        \input{heat_map_shapes_tikz.tex}
    }
    \caption{Heat map radiation patterns for different activation shapes of a $16\times16$ shape-adaptive RHS.}
    \label{fig:heat_map}
\end{figure*}

The fundamental motivation for shape adaptation is the strong dependence of radiation characteristics on the geometry of the active surface. Different RHS shapes produce distinct radiation patterns, beamwidths, and side-lobe behaviors, even when operating at the same carrier frequency and total radiated power. 
Fig.~\ref{fig:activation_patterns} illustrates different shapes for an RHS, while Fig.~\ref{fig:heat_map} depicts heat maps of the obtained radiation patterns. 
Based on these figures, a configuration with a rectangular region yields a highly directive main lobe with a relatively narrow beamwidth, making it well-suited for serving users within well-defined angular sectors. 
In contrast, a column-wise or striped activation pattern produces a wider beam with increased angular coverage at the expense of peak gain. 

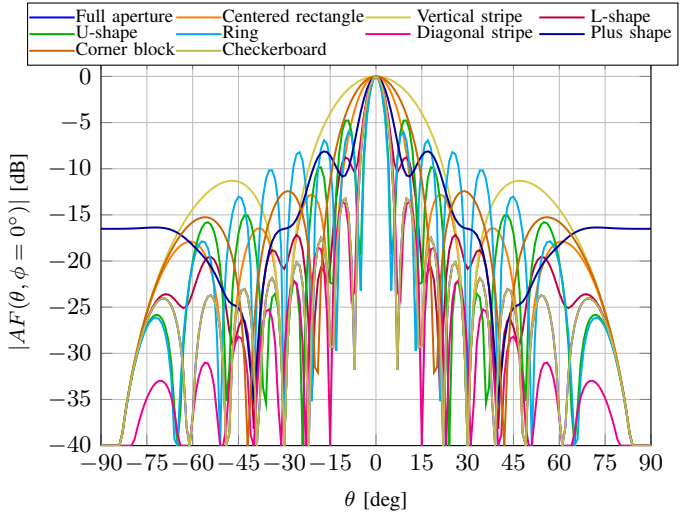
\begin{figure}[t]
    \centering
    \input{elevation_cut}
    \vspace{-2.5mm}
    \caption{Elevation cut ($\phi=0^\circ$) of the normalized array factor for different activation shapes of a $16\times16$ shape-adaptive RHS. Each activation pattern produces a distinct radiation pattern, illustrating how the effective aperture geometry influences beamwidth, sidelobe levels, and angular selectivity.}
    \label{fig:elevation_cut}
\end{figure}

Fig.~\ref{fig:elevation_cut} shows the elevation cut of radiation patterns for various RHS shapes. This figure demonstrates that irregular or non-rectangular shapes, e.g., L-shaped or diagonally activated regions, can generate asymmetric radiation patterns. 
These patterns can be exploited to mitigate blockages, avoid strong interferers, or steer energy around obstacles. 
The ability to switch among these shapes enables RHSs to adapt their spatial radiation characteristics without modifying the physical deployment or hardware. 
These results highlight that \textit{shape adaptation provides a form of beam control that cannot be achieved through phase tuning alone}. 
By modifying the radiating aperture, RHSs can tailor their angular spectrum to better match the propagation environment and user distribution.

\section{Shape-Adaptive RHS-Aided Communication}
Unlike conventional RIS- or RHS-assisted systems with fixed surface geometries~\cite{9696209}, \textcolor{black}{shape-adaptive RHSs introduce an additional optimization variable into wireless communication systems: dynamic shape selection. 
This allows the aperture configuration to be jointly designed with conventional resource allocation policies, e.g., transmit beamforming, power allocation, and user scheduling. 
This additional flexibility, i.e., shape adaptivity, can conceptually extend beyond static infrastructure-based communications to emerging scenarios, e.g., UAV-assisted networks and integrated sensing and communication (ISAC), where different propagation environments may favor different radiation patterns.} 
In the following, we focus on a downlink communication scenario as a representative use case to illustrate the proposed framework and demonstrate the performance gains enabled by adaptive aperture shaping.

\subsection{An Exemplary Shape-adaptive RHS-aided System Model}

Consider a downlink communication system in which a multi-antenna BS serves multiple user equipments (UEs) with the assistance of several distributed shape-adaptive RHSs. Prior to each transmission interval, the network controller determines the BS beamformer, the RHS radiation coefficients, and one aperture configuration for every deployed RHS according to the available channel state information (CSI), user locations, or other network objectives. These configuration parameters are then communicated to the corresponding RHS controllers through low-rate control links.

{\color{black}
During transmission, the BS first generates the multi-user baseband signal by precoding and superimposing the data streams intended for different UEs. The corresponding RF signal is delivered to the embedded feed of each assisting RHS through an appropriate feeder interface. The embedded feed excites a guided reference wave that propagates across the holographic surface. According to the selected aperture configuration, only the active meta-elements extract energy from the guided wave and radiate it into free space, while their radiation coefficients determine the amplitude contribution of each active element to the synthesized wavefront.

The received signal at each UE consists of the superposition of the direct BS transmission, the signals radiated by all assisting shape-adaptive RHSs, and receiver noise. The combined signal quality therefore depends on the joint design of the BS active beamformer, the selected aperture configuration, and the radiation coefficients, which together determine the spatial distribution of the transmitted energy. 
This joint optimization allows the communication system to adapt its propagation characteristics to changing network conditions while maintaining compatibility with conventional multi-user downlink transmission. The resulting signal model provides the basis for the numerical performance evaluation presented in Subsection~\ref{results}.}

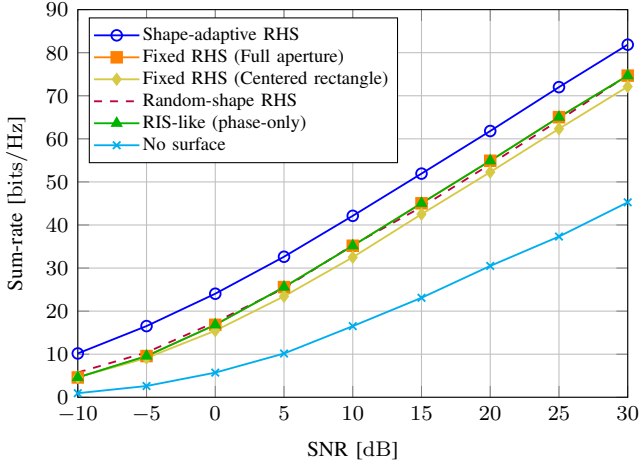
\begin{figure}[t]
    \centering
    \input{sum_rate_vs_snr}
    \caption{Downlink sum-rate versus SNR for the proposed shape-adaptive RHS compared with other baselines.}
    \label{fig:sum_rate}
\end{figure}

\subsection{Joint Optimization of Shape, Beamforming and Parameters}

To fully exploit the benefits of shape adaptation, a joint optimization problem can be adopted that integrates three interacting optimization variables: (i) active beamforming at the BS, (ii) surface parameter control, i.e., passive beamforming, at the RHS, and (iii) discrete shape selection for each RHS.  The shape selection problem is inherently combinatorial, as each RHS chooses one configuration from a \textit{finite library of candidate shapes}.  
\textcolor{black}{In a distributed deployment, an exhaustive search over all RHS shape combinations scales with the product of the per-RHS library sizes, e.g., \(K^R\) for \(R\) RHSs with \(K\) candidate shapes each; optimizing the binary activation state of every element would be even more computationally prohibitive. Therefore, the shape library is intended as a compact codebook of coverage-relevant aperture masks rather than an enumeration of all possible element states.}  To limit the computational cost, shape adaptation is treated as a discrete optimization layer that operates on a slower timescale than phase or amplitude tuning.  \textcolor{black}{Candidate shapes can also be ranked or pruned using long-term channel statistics, user angular sectors, blockage maps, or previously observed utility values before the online alternating optimization (AO) is performed.  For a given shape configuration, the BS beamformer and RHS surface parameters are optimized to maximize a network-level performance metric, e.g., the sum-rate or error rate.}
An AO algorithm can be employed, where the BS beamformer and RHS parameters are updated assuming a fixed shape, followed by shape selection based on a performance criterion.  This approach balances performance gains with implementation cost and enables high-performance operation without requiring an exhaustive search over all possible configurations.

\textcolor{black}{Power leakage constraints are naturally incorporated into the above optimization problem to ensure physically realizable shape-adaptive RHS configurations~\cite{10163760}. 
Specifically, for each candidate shape, the total leaked power is computed by aggregating the guided power extracted and radiated by all active meta-elements according to their controllable radiation coefficients.  The resulting leakage is constrained to remain within a prescribed operating interval, where the lower bound prevents excessive confinement of the guided wave and insufficient radiation, while the upper bound avoids premature surface-wave dissipation, unintended radiation, and violation of the available power budget. In addition, the radiation coefficient of each active meta-element is constrained by its hardware-supported operating range, whereas inactive elements are assigned zero radiation coefficients. Consequently, every AO update of the BS beamformer, RHS radiation coefficients, and aperture configuration must satisfy these feasibility constraints. Alternatively, when strict feasibility is not required, the leakage constraints defining the feasible region can be relaxed and incorporated into the objective function as a soft penalty, allowing a controlled trade-off between communication performance and deviation from the desired leakage level.}

\subsection{Multi-User Throughput}\label{results}
\textcolor{black}{The simulations consider $6$ single-antenna UEs, a BS equipped with $16$ antennas, and a shape-adaptive RHS comprising $256$ elements arranged in $16$ rows and $16$ columns. The direct BS--UE channels follow independent Rayleigh fading, while the BS--RHS and RHS--UE channels follow Rician fading with linear Rician factors of 5 and 3, respectively.  All average channel-power gains are normalized to unity. Each direct BS--UE link is independently blocked with probability $0.35$, in which case its average channel-power gain is reduced by $30$ $\rm{dB}$. The receiver noise power is obtained from the standard thermal-noise model using a temperature of $290$ Kelvin, a bandwidth of $100$ $\rm{MHz}$, and a receiver noise figure of $7$ $\rm{dB}$, resulting in a noise power of $-87$ $\rm{dBm}$. The transmit power is varied so that the ratio of transmit power to noise power ranges from $-10$ to $30$ $\rm{dB}$. Equal power allocation and zero-forcing (ZF) beamforming are employed, and the sum-rates are obtained by averaging over $200$ independent Monte Carlo channel realizations.}

Fig.~\ref{fig:sum_rate} shows the multi-user throughput benefits of utilizing shape adaptation into RHS-assisted downlink systems. 
By dynamically selecting activation patterns that best match the users’ angular distribution and prevailing blockage conditions, the proposed approach improves spatial separation among users and effectively suppresses inter-user interference. 
This advantage becomes increasingly pronounced at moderate-to-high SNR, where interference rather than noise limits performance. 
In contrast, fixed-geometry, random-shape, and RIS-like phase-only baselines provide aperture gain but cannot align the effective radiation geometry with the network topology, leading to a noticeable loss in sum-rate. 
This indicates that irregular, shape-adaptive configurations can allow robust connectivity in blockage-dominated scenarios by enabling alternative propagation paths, supporting more efficient spatial multiplexing without increasing transmit power or hardware complexity.

\section{Research Challenges and Future Directions}
While shape-adaptive RHSs offer powerful new capabilities for wireless communication and sensing, several research and engineering challenges must be addressed to enable their large-scale deployment. This section highlights key open problems and outlines promising directions for future investigation.
\vspace{-4mm}
\subsection{Real-Time Shape Control under Hardware Constraints}
One of the primary challenges lies in realizing real-time shape adaptation under practical hardware constraints. Dynamic reconfiguration of RHS requires coordinated control of a large number of meta-elements, often with stringent latency and energy-efficiency requirements. Controllers must be capable of rapidly switching between shape configurations while respecting power leakage constraints, hardware non-idealities, and EM compatibility regulations. \textcolor{black}{In practical deployments, the RHS controller will generally rely on estimated or partial CSI and finite-resolution radiation/phase states, while shape updates must be scheduled to avoid excessive pilot overhead and hardware-response delays. Designing robust, low-overhead channel-estimation and control protocols under these non-idealities is an important direction for future work.} Moreover, shape adaptation introduces an extra control dimension that operates alongside phase and amplitude tuning. Efficient scheduling of these control actions across different timescales remains an open problem. Developing lightweight, low-latency control architectures that can adapt RHSs' shapes in response to dynamic channel and network conditions is essential for real-time operation in dense and mobile scenarios.

\vspace{-4mm}
\subsection{Joint Control Across Multiple RHS Surfaces}
In realistic deployments, RHS units may be distributed across a wide geographic area, jointly assisting communication and sensing tasks. Coordinating shape adaptation across these surfaces introduces scalability challenges. 
Centralized optimization quickly becomes infeasible as the number of RHSs grows, motivating the need for distributed and hierarchical control architectures. 
Future research must address how multiple RHSs can cooperatively select shapes and surface parameters while accounting for mutual interference, signaling overhead, and partial channel state information. 
Consensus-based algorithms, game-theoretic formulations, and federated optimization approaches are promising tools for enabling scalable and robust coordination of RHSs in large networks.

\vspace{-3mm}

\subsection{Machine Learning for Adaptive Surface Shaping}
{\color{black} 
The optimization framework and numerical results in this paper are based on a model-driven alternating optimization algorithm. Accordingly, the benchmark comparisons are designed to isolate the architectural benefits of adaptive shaping relative to fixed-geometry, random-shape, and phase-only configurations. Learning-based methods~\cite{11008469}, e.g., deep reinforcement learning (DRL) and graph neural networks (GNNs), are not considered as algorithmic baselines in this study; 
instead, they are envisioned as promising future extensions for scalable, real-time control in large and dynamic RHS deployments. 
In particular, DRL could learn shape-selection policies from network traffic observations, while GNNs could capture the interactions among users, RHSs, and access points in distributed deployments. 
A key challenge is achieving fast and reliable adaptation with limited training data, partial CSI, and stringent latency constraints.}

\vspace{-3mm}
\subsection{Standardization and Measurement Campaigns}
The transition of shape-adaptive RHSs from research prototypes to commercial systems requires standardization and experimental validation. 
A unified framework for RHS architectures, control interfaces, and performance metrics is needed to compare results and support integration into communication standards. 
Large-scale measurement campaigns and testbeds should validate theoretical models, characterize real-world propagation effects, and define reference scenarios, calibration procedures, and interoperability guidelines.

\section{Conclusion}
In this work, we proposed a novel shape-adaptive deployment strategy for RHSs that addresses key limitations of existing RIS optimization approaches. Unlike conventional methods based on RIS selection, placement, or binary element activation, the proposed framework introduces \textcolor{black}{additional spatial degrees of freedom by dynamically selecting and configuring the geometric shape of the effective RHS aperture.
This enables the wireless system to better adapt to changing propagation environments, user mobility, and blockage conditions while improving beamforming flexibility and spatial resource utilization. We also discussed the fundamental principles, practical implementation aspects, representative use cases, and performance benefits of shape-adaptive RHSs, demonstrating their potential for future high-frequency wireless networks.}
Continued research is needed on efficient optimization, scalable distributed control architectures, and hardware prototyping to facilitate practical deployment in next-generation communication and sensing networks.

\bibliographystyle{ieeetr}
\bibliography{ref}
\end{document}

%% file: rhs_shapes.tex
\begin{tikzpicture}

\newcommand{\drawshape}[3]{%
\begin{scope}[shift={#1}, scale=0.20]
\node[font=\bfseries\footnotesize] at (8.5,17.3) {#2};

\foreach \x in {1,...,16}{
\foreach \y in {1,...,16}{
    \pgfmathtruncatemacro{\on}{#3}
    \ifnum\on=1
        \fill[white] (\x-1,16-\y) rectangle (\x,17-\y);
    \else
        \fill[black] (\x-1,16-\y) rectangle (\x,17-\y);
    \fi
}}
\draw[black, thin] (0,0) rectangle (16,16);

\foreach \t in {5,10,15}{
    \draw (\t,0) -- ++(0,-0.25) node[below,font=\scriptsize] {\t};
    \draw (0,16-\t+1) -- ++(-0.25,0) node[left,font=\scriptsize] {\t};
}

\node[font=\scriptsize] at (8,-2.4) {$x$ index};
\node[font=\scriptsize, rotate=90] at (-3.3,8) {$y$ index};
\end{scope}
}


\drawshape{(0,0)}{Full aperture}{0}
\drawshape{(4.2,0)}{Centered rectangle}{(\x>=5 && \x<=12 && \y>=5 && \y<=12)}
\drawshape{(8.4,0)}{Vertical stripe}{(\x>=7 && \x<=10)}
\drawshape{(12.6,0)}{L-shape}{(\x<=5 || \y>=12)}
\drawshape{(16.8,0)}{U-shape}{(\x>=5 && \x<=12 && \y<=12)}

\drawshape{(0,-4.3)}{Ring}{(\x<=2 || \x>=15 || \y<=2 || \y>=15)}
\drawshape{(4.2,-4.3)}{Diagonal stripe}{abs(\x-\y)<=1}
\drawshape{(8.4,-4.3)}{Plus shape}{(\x>=8 && \x<=10) || (\y>=8 && \y<=10)}
\drawshape{(12.6,-4.3)}{Corner block}{(\x<=6 && \y>=11)}
\drawshape{(16.8,-4.3)}{Checkerboard}{mod(\x+\y,2)==0}

\end{tikzpicture}

%% file: heat_map_shapes_tikz.tex

\pgfplotsset{
colormap={parulaish}{
    rgb255=(53,42,135)
    rgb255=(15,92,221)
    rgb255=(18,125,216)
    rgb255=(7,156,207)
    rgb255=(21,177,180)
    rgb255=(89,189,140)
    rgb255=(165,190,107)
    rgb255=(225,185,82)
    rgb255=(252,206,46)
    rgb255=(249,251,14)
}}
\begin{tikzpicture}
\begin{groupplot}[
    group style={group size=5 by 2, horizontal sep=1.0cm, vertical sep=1.0cm},
    width=0.22\textwidth, height=0.30\textwidth,
    ylabel={$\phi$ \rm{[deg]}},
    xmin=-90, xmax=90, ymin=-180, ymax=180,
    xtick={-90,-45,0,45,90}, ytick={-180,-135,-90,-45,0,45,90,135,180},
    point meta min=-30, point meta max=0,
    colormap name=parulaish,
    tick label style={font=\scriptsize}, label style={font=\scriptsize}, title style={font=\scriptsize\bfseries},
    view={0}{90}, axis on top
]
\nextgroupplot[title={Full aperture}]
\addplot3[surf, shader=interp, mesh/rows=31, mesh/cols=31, draw=none] table[x=theta,y=phi,z=afdb] {shape_1.dat};
\nextgroupplot[title={Centered rectangle}]
\addplot3[surf, shader=interp, mesh/rows=31, mesh/cols=31, draw=none] table[x=theta,y=phi,z=afdb] {shape_2.dat};
\nextgroupplot[title={Vertical stripe}]
\addplot3[surf, shader=interp, mesh/rows=31, mesh/cols=31, draw=none] table[x=theta,y=phi,z=afdb] {shape_3.dat};
\nextgroupplot[title={L-shape}]
\addplot3[surf, shader=interp, mesh/rows=31, mesh/cols=31, draw=none] table[x=theta,y=phi,z=afdb] {shape_4.dat};
\nextgroupplot[
    title={U-shape},
    colorbar,
    colorbar style={
        width=1.2mm,
        ylabel={$|AF(\theta,\phi)|$ \rm{[dB]}},
        ytick={-30,-25,-20,-15,-10,-5,0},
        yticklabel style={font=\tiny},
        ylabel style={font=\tiny}
    }
]
\addplot3[surf, shader=interp, mesh/rows=31, mesh/cols=31, draw=none] table[x=theta,y=phi,z=afdb] {shape_5.dat};
\nextgroupplot[title={Ring}]
\addplot3[surf, shader=interp, mesh/rows=31, mesh/cols=31, draw=none] table[x=theta,y=phi,z=afdb] {shape_6.dat};
\nextgroupplot[title={Diagonal stripe}]
\addplot3[surf, shader=interp, mesh/rows=31, mesh/cols=31, draw=none] table[x=theta,y=phi,z=afdb] {shape_7.dat};
\nextgroupplot[title={Plus shape},xlabel={$\theta$ \rm{[deg]}}]
\addplot3[surf, shader=interp, mesh/rows=31, mesh/cols=31, draw=none] table[x=theta,y=phi,z=afdb] {shape_8.dat};
\nextgroupplot[title={Corner block}]
\addplot3[surf, shader=interp, mesh/rows=31, mesh/cols=31, draw=none] table[x=theta,y=phi,z=afdb] {shape_9.dat};
\nextgroupplot[
    title={Checkerboard},
    xlabel={$\theta$ (deg)},
    colorbar,
    colorbar style={
        width=1.2mm,
        ylabel={$|AF(\theta,\phi)|$ \rm{[dB]}},
        ytick={-30,-25,-20,-15,-10,-5,0},
        yticklabel style={font=\tiny},
        ylabel style={font=\tiny}
    }
]
\addplot3[surf, shader=interp, mesh/rows=31, mesh/cols=31, draw=none] table[x=theta,y=phi,z=afdb] {shape_10.dat};
\end{groupplot}
\end{tikzpicture}

%% file: elevation_cut.tex
\begin{tikzpicture}
\begin{axis}[
    width=1\columnwidth,
    height=0.37\textwidth,
    xlabel={$\theta$ [\rm{deg}]},
    ylabel={$|AF(\theta,\phi=0^\circ)|$ [\rm{dB}]},
    grid=major,
    legend style={
        at={(1.05,0.98)},
        anchor=south east,
        font=\scriptsize,
        inner sep=+0.1mm,
        legend cell align={left},
        legend columns=4,
        /tikz/column 1/.style={column sep=-2pt,},
        /tikz/column 2/.style={column sep=-2pt,},
        /tikz/column 3/.style={column sep=-2pt,},
        /tikz/column 4/.style={column sep=-2pt,},
        /tikz/row 1/.style={row sep=-4pt,},
        /tikz/row 2/.style={row sep=-4pt,},
        /tikz/row 3/.style={row sep=-5pt,},
    },
    tick label style={font=\footnotesize},
    xlabel style={font=\footnotesize},
    ylabel style={font=\footnotesize},
    xmin=-90, xmax=90,
    ymin=-40, ymax=2.0,
    xtick={-90,-75,-60,-45,-30,-15,0,15,30,45,60,75,90},
    ytick={-40,-35,-30,-25,-20,-15,-10,-5,0},
    tick label style={font=\small},
    xlabel style={font=\footnotesize},
    ylabel style={font=\footnotesize},
    title style={font=\footnotesize},
    clip=true
]
\addplot[solid, blue, line width=0.75pt] coordinates {
    (-90,-40.000000) (-89,-40.000000) (-88,-40.000000) (-87,-40.000000) (-86,-40.000000) (-85,-40.000000)
    (-84,-40.000000) (-83,-38.680260) (-82,-36.399385) (-81,-34.408695) (-80,-32.654775) (-79,-31.101364)
    (-78,-29.723636) (-77,-28.504818) (-76,-27.434135) (-75,-26.505561) (-74,-25.717088) (-73,-25.070407)
    (-72,-24.570923) (-71,-24.228145) (-70,-24.056563) (-69,-24.077253) (-68,-24.320732) (-67,-24.832183)
    (-66,-25.681615) (-65,-26.985696) (-64,-28.962030) (-63,-32.099294) (-62,-37.983071) (-61,-40.000000)
    (-60,-36.897303) (-59,-31.125547) (-58,-27.890708) (-57,-25.832792) (-56,-24.538263) (-55,-23.849069)
    (-54,-23.718861) (-53,-24.178802) (-52,-25.352592) (-51,-27.544681) (-50,-31.607562) (-49,-40.000000)
    (-48,-38.673220) (-47,-30.188565) (-46,-26.350448) (-45,-24.228981) (-44,-23.183832) (-43,-23.033754)
    (-42,-23.812881) (-41,-25.800025) (-40,-29.920766) (-39,-40.000000) (-38,-35.062361) (-37,-27.447783)
    (-36,-24.004436) (-35,-22.308952) (-34,-21.838917) (-33,-22.546510) (-32,-24.772440) (-31,-29.931644)
    (-30,-40.000000) (-29,-29.435684) (-28,-23.805641) (-27,-21.165009) (-26,-20.146116) (-25,-20.498403)
    (-24,-22.497241) (-23,-27.577497) (-22,-40.000000) (-21,-26.458427) (-20,-20.912533) (-19,-18.352829)
    (-18,-17.493958) (-17,-18.178329) (-16,-20.965764) (-15,-29.182123) (-14,-29.377000) (-13,-19.465988)
    (-12,-15.404311) (-11,-13.516407) (-10,-13.227569) (-9,-14.774075) (-8,-19.956980) (-7,-31.761029)
    (-6,-14.509275) (-5,-8.570309) (-4,-5.004357) (-3,-2.657550) (-2,-1.139424) (-1,-0.279238)
    (0,0.000000) (1,-0.279238) (2,-1.139424) (3,-2.657550) (4,-5.004357) (5,-8.570309)
    (6,-14.509275) (7,-31.761029) (8,-19.956980) (9,-14.774075) (10,-13.227569) (11,-13.516407)
    (12,-15.404311) (13,-19.465988) (14,-29.377000) (15,-29.182123) (16,-20.965764) (17,-18.178329)
    (18,-17.493958) (19,-18.352829) (20,-20.912533) (21,-26.458427) (22,-40.000000) (23,-27.577497)
    (24,-22.497241) (25,-20.498403) (26,-20.146116) (27,-21.165009) (28,-23.805641) (29,-29.435684)
    (30,-40.000000) (31,-29.931644) (32,-24.772440) (33,-22.546510) (34,-21.838917) (35,-22.308952)
    (36,-24.004436) (37,-27.447783) (38,-35.062361) (39,-40.000000) (40,-29.920766) (41,-25.800025)
    (42,-23.812881) (43,-23.033754) (44,-23.183832) (45,-24.228981) (46,-26.350448) (47,-30.188565)
    (48,-38.673220) (49,-40.000000) (50,-31.607562) (51,-27.544681) (52,-25.352592) (53,-24.178802)
    (54,-23.718861) (55,-23.849069) (56,-24.538263) (57,-25.832792) (58,-27.890708) (59,-31.125547)
    (60,-36.897303) (61,-40.000000) (62,-37.983071) (63,-32.099294) (64,-28.962030) (65,-26.985696)
    (66,-25.681615) (67,-24.832183) (68,-24.320732) (69,-24.077253) (70,-24.056563) (71,-24.228145)
    (72,-24.570923) (73,-25.070407) (74,-25.717088) (75,-26.505561) (76,-27.434135) (77,-28.504818)
    (78,-29.723636) (79,-31.101364) (80,-32.654775) (81,-34.408695) (82,-36.399385) (83,-38.680260)
    (84,-40.000000) (85,-40.000000) (86,-40.000000) (87,-40.000000) (88,-40.000000) (89,-40.000000)
    (90,-40.000000)
};
\addlegendentry{Full aperture}

\addplot[solid, orange, line width=0.75pt] coordinates {
    (-90,-40.000000) (-89,-40.000000) (-88,-40.000000) (-87,-40.000000) (-86,-40.000000) (-85,-40.000000)
    (-84,-40.000000) (-83,-38.642101) (-82,-36.334269) (-81,-34.304325) (-80,-32.495516) (-79,-30.867775)
    (-78,-29.391942) (-77,-28.046302) (-76,-26.814417) (-75,-25.683709) (-74,-24.644506) (-73,-23.689383)
    (-72,-22.812697) (-71,-22.010263) (-70,-21.279112) (-69,-20.617331) (-68,-20.023955) (-67,-19.498898)
    (-66,-19.042932) (-65,-18.657701) (-64,-18.345780) (-63,-18.110782) (-62,-17.957536) (-61,-17.892344)
    (-60,-17.923378) (-59,-18.061267) (-58,-18.319996) (-57,-18.718320) (-56,-19.282097) (-55,-20.048313)
    (-54,-21.072557) (-53,-22.444227) (-52,-24.321579) (-51,-27.029189) (-50,-31.429809) (-49,-40.000000)
    (-48,-38.640951) (-47,-29.947905) (-46,-25.689142) (-45,-22.900915) (-44,-20.888451) (-43,-19.379378)
    (-42,-18.242014) (-41,-17.406324) (-40,-16.834864) (-39,-16.510633) (-38,-16.432178) (-37,-16.612880)
    (-36,-17.083993) (-35,-17.903233) (-34,-19.174985) (-33,-21.101612) (-32,-24.143495) (-31,-29.775620)
    (-30,-40.000000) (-29,-29.276464) (-28,-23.150154) (-27,-19.624320) (-26,-17.229868) (-25,-15.513381)
    (-24,-14.281173) (-23,-13.440246) (-22,-12.948647) (-21,-12.798211) (-20,-13.011621) (-19,-13.651202)
    (-18,-14.847654) (-17,-16.884055) (-16,-20.506978) (-15,-29.128674) (-14,-29.332170) (-13,-19.028377)
    (-12,-14.128190) (-11,-10.852666) (-10,-8.405171) (-9,-6.479319) (-8,-4.924506) (-7,-3.655694)
    (-6,-2.620465) (-5,-1.784598) (-4,-1.124956) (-3,-0.625655) (-2,-0.275890) (-1,-0.068654)
    (0,0.000000) (1,-0.068654) (2,-0.275890) (3,-0.625655) (4,-1.124956) (5,-1.784598)
    (6,-2.620465) (7,-3.655694) (8,-4.924506) (9,-6.479319) (10,-8.405171) (11,-10.852666)
    (12,-14.128190) (13,-19.028377) (14,-29.332170) (15,-29.128674) (16,-20.506978) (17,-16.884055)
    (18,-14.847654) (19,-13.651202) (20,-13.011621) (21,-12.798211) (22,-12.948647) (23,-13.440246)
    (24,-14.281173) (25,-15.513381) (26,-17.229868) (27,-19.624320) (28,-23.150154) (29,-29.276464)
    (30,-40.000000) (31,-29.775620) (32,-24.143495) (33,-21.101612) (34,-19.174985) (35,-17.903233)
    (36,-17.083993) (37,-16.612880) (38,-16.432178) (39,-16.510633) (40,-16.834864) (41,-17.406324)
    (42,-18.242014) (43,-19.379378) (44,-20.888451) (45,-22.900915) (46,-25.689142) (47,-29.947905)
    (48,-38.640951) (49,-40.000000) (50,-31.429809) (51,-27.029189) (52,-24.321579) (53,-22.444227)
    (54,-21.072557) (55,-20.048313) (56,-19.282097) (57,-18.718320) (58,-18.319996) (59,-18.061267)
    (60,-17.923378) (61,-17.892344) (62,-17.957536) (63,-18.110782) (64,-18.345780) (65,-18.657701)
    (66,-19.042932) (67,-19.498898) (68,-20.023955) (69,-20.617331) (70,-21.279112) (71,-22.010263)
    (72,-22.812697) (73,-23.689383) (74,-24.644506) (75,-25.683709) (76,-26.814417) (77,-28.046302)
    (78,-29.391942) (79,-30.867775) (80,-32.495516) (81,-34.304325) (82,-36.334269) (83,-38.642101)
    (84,-40.000000) (85,-40.000000) (86,-40.000000) (87,-40.000000) (88,-40.000000) (89,-40.000000)
    (90,-40.000000)
};
\addlegendentry{Centered rectangle}

\addplot[solid, yellow!80!black, line width=0.75pt] coordinates {
    (-90,-40.000000) (-89,-40.000000) (-88,-40.000000) (-87,-40.000000) (-86,-40.000000) (-85,-40.000000)
    (-84,-40.000000) (-83,-38.632572) (-82,-36.318020) (-81,-34.278311) (-80,-32.455883) (-79,-30.809770)
    (-78,-29.309810) (-77,-27.933186) (-76,-26.662251) (-75,-25.483105) (-74,-24.384634) (-73,-23.357836)
    (-72,-22.395347) (-71,-21.491092) (-70,-20.640022) (-69,-19.837926) (-68,-19.081275) (-67,-18.367111)
    (-66,-17.692955) (-65,-17.056735) (-64,-16.456733) (-63,-15.891536) (-62,-15.360004) (-61,-14.861240)
    (-60,-14.394570) (-59,-13.959527) (-58,-13.555837) (-57,-13.183417) (-56,-12.842367) (-55,-12.532974)
    (-54,-12.255719) (-53,-12.011285) (-52,-11.800576) (-51,-11.624735) (-50,-11.485182) (-49,-11.383648)
    (-48,-11.322233) (-47,-11.303476) (-46,-11.330448) (-45,-11.406881) (-44,-11.537340) (-43,-11.727457)
    (-42,-11.984266) (-41,-12.316681) (-40,-12.736208) (-39,-13.258028) (-38,-13.902730) (-37,-14.699198)
    (-36,-15.689750) (-35,-16.940000) (-34,-18.559867) (-33,-20.755392) (-32,-23.989105) (-31,-29.736790)
    (-30,-40.000000) (-29,-29.236841) (-28,-22.989373) (-27,-19.256205) (-26,-16.561716) (-25,-14.443537)
    (-24,-12.695787) (-23,-11.208703) (-22,-9.916823) (-21,-8.777979) (-20,-7.763430) (-19,-6.852725)
    (-18,-6.030816) (-17,-5.286321) (-16,-4.610444) (-15,-3.996257) (-14,-3.438227) (-13,-2.931874)
    (-12,-2.473541) (-11,-2.060213) (-10,-1.689396) (-9,-1.359012) (-8,-1.067337) (-7,-0.812933)
    (-6,-0.594612) (-5,-0.411399) (-4,-0.262509) (-3,-0.147321) (-2,-0.065369) (-1,-0.016326)
    (0,0.000000) (1,-0.016326) (2,-0.065369) (3,-0.147321) (4,-0.262509) (5,-0.411399)
    (6,-0.594612) (7,-0.812933) (8,-1.067337) (9,-1.359012) (10,-1.689396) (11,-2.060213)
    (12,-2.473541) (13,-2.931874) (14,-3.438227) (15,-3.996257) (16,-4.610444) (17,-5.286321)
    (18,-6.030816) (19,-6.852725) (20,-7.763430) (21,-8.777979) (22,-9.916823) (23,-11.208703)
    (24,-12.695787) (25,-14.443537) (26,-16.561716) (27,-19.256205) (28,-22.989373) (29,-29.236841)
    (30,-40.000000) (31,-29.736790) (32,-23.989105) (33,-20.755392) (34,-18.559867) (35,-16.940000)
    (36,-15.689750) (37,-14.699198) (38,-13.902730) (39,-13.258028) (40,-12.736208) (41,-12.316681)
    (42,-11.984266) (43,-11.727457) (44,-11.537340) (45,-11.406881) (46,-11.330448) (47,-11.303476)
    (48,-11.322233) (49,-11.383648) (50,-11.485182) (51,-11.624735) (52,-11.800576) (53,-12.011285)
    (54,-12.255719) (55,-12.532974) (56,-12.842367) (57,-13.183417) (58,-13.555837) (59,-13.959527)
    (60,-14.394570) (61,-14.861240) (62,-15.360004) (63,-15.891536) (64,-16.456733) (65,-17.056735)
    (66,-17.692955) (67,-18.367111) (68,-19.081275) (69,-19.837926) (70,-20.640022) (71,-21.491092)
    (72,-22.395347) (73,-23.357836) (74,-24.384634) (75,-25.483105) (76,-26.662251) (77,-27.933186)
    (78,-29.309810) (79,-30.809770) (80,-32.455883) (81,-34.278311) (82,-36.318020) (83,-38.632572)
    (84,-40.000000) (85,-40.000000) (86,-40.000000) (87,-40.000000) (88,-40.000000) (89,-40.000000)
    (90,-40.000000)
};
\addlegendentry{Vertical stripe}

\addplot[solid, purple, line width=0.75pt] coordinates {
    (-90,-40.000000) (-89,-40.000000) (-88,-40.000000) (-87,-40.000000) (-86,-40.000000) (-85,-40.000000)
    (-84,-40.000000) (-83,-38.677534) (-82,-36.394707) (-81,-34.401136) (-80,-32.643108) (-79,-31.083988)
    (-78,-29.698460) (-77,-28.469101) (-76,-27.384248) (-75,-26.436635) (-74,-25.622478) (-73,-24.940862)
    (-72,-24.393270) (-71,-23.983150) (-70,-23.715340) (-69,-23.595054) (-68,-23.625830) (-67,-23.805278)
    (-66,-24.116523) (-65,-24.512474) (-64,-24.893301) (-63,-25.094868) (-62,-24.936828) (-61,-24.346786)
    (-60,-23.431533) (-59,-22.390791) (-58,-21.398774) (-57,-20.566523) (-56,-19.957612) (-55,-19.611432)
    (-54,-19.559579) (-53,-19.836039) (-52,-20.483994) (-51,-21.560058) (-50,-23.129736) (-49,-25.217824)
    (-48,-27.560165) (-47,-29.007132) (-46,-28.480482) (-45,-27.180224) (-44,-26.365679) (-43,-26.481048)
    (-42,-27.850586) (-41,-31.238435) (-40,-38.185822) (-39,-33.111743) (-38,-26.628630) (-37,-22.916745)
    (-36,-20.650139) (-35,-19.351936) (-34,-18.823387) (-33,-18.960249) (-32,-19.636867) (-31,-20.515435)
    (-30,-20.852192) (-29,-20.093009) (-28,-18.767122) (-27,-17.661917) (-26,-17.171107) (-25,-17.484542)
    (-24,-18.817514) (-23,-21.634459) (-22,-27.296546) (-21,-34.680544) (-20,-26.058464) (-19,-22.088451)
    (-18,-20.735780) (-17,-21.114087) (-16,-22.197238) (-15,-20.593347) (-14,-16.581353) (-13,-13.155713)
    (-12,-10.786671) (-11,-9.357889) (-10,-8.770200) (-9,-8.959800) (-8,-9.753063) (-7,-10.389913)
    (-6,-9.399107) (-5,-6.935755) (-4,-4.421339) (-3,-2.441011) (-2,-1.066390) (-1,-0.263648)
    (0,0.000000) (1,-0.263648) (2,-1.066390) (3,-2.441011) (4,-4.421339) (5,-6.935755)
    (6,-9.399107) (7,-10.389913) (8,-9.753063) (9,-8.959800) (10,-8.770200) (11,-9.357889)
    (12,-10.786671) (13,-13.155713) (14,-16.581353) (15,-20.593347) (16,-22.197238) (17,-21.114087)
    (18,-20.735780) (19,-22.088451) (20,-26.058464) (21,-34.680544) (22,-27.296546) (23,-21.634459)
    (24,-18.817514) (25,-17.484542) (26,-17.171107) (27,-17.661917) (28,-18.767122) (29,-20.093009)
    (30,-20.852192) (31,-20.515435) (32,-19.636867) (33,-18.960249) (34,-18.823387) (35,-19.351936)
    (36,-20.650139) (37,-22.916745) (38,-26.628630) (39,-33.111743) (40,-38.185822) (41,-31.238435)
    (42,-27.850586) (43,-26.481048) (44,-26.365679) (45,-27.180224) (46,-28.480482) (47,-29.007132)
    (48,-27.560165) (49,-25.217824) (50,-23.129736) (51,-21.560058) (52,-20.483994) (53,-19.836039)
    (54,-19.559579) (55,-19.611432) (56,-19.957612) (57,-20.566523) (58,-21.398774) (59,-22.390791)
    (60,-23.431533) (61,-24.346786) (62,-24.936828) (63,-25.094868) (64,-24.893301) (65,-24.512474)
    (66,-24.116523) (67,-23.805278) (68,-23.625830) (69,-23.595054) (70,-23.715340) (71,-23.983150)
    (72,-24.393270) (73,-24.940862) (74,-25.622478) (75,-26.436635) (76,-27.384248) (77,-28.469101)
    (78,-29.698460) (79,-31.083988) (80,-32.643108) (81,-34.401136) (82,-36.394707) (83,-38.677534)
    (84,-40.000000) (85,-40.000000) (86,-40.000000) (87,-40.000000) (88,-40.000000) (89,-40.000000)
    (90,-40.000000)
};
\addlegendentry{L-shape}

\addplot[solid, green!70!black, line width=0.75pt] coordinates {
    (-90,-40.000000) (-89,-40.000000) (-88,-40.000000) (-87,-40.000000) (-86,-40.000000) (-85,-40.000000)
    (-84,-40.000000) (-83,-38.703237) (-82,-36.438690) (-81,-34.471925) (-80,-32.751751) (-79,-31.244593)
    (-78,-29.928908) (-77,-28.792014) (-76,-27.828365) (-75,-27.038792) (-74,-26.430577) (-73,-26.018435)
    (-72,-25.826782) (-71,-25.894252) (-70,-26.282678) (-69,-27.096345) (-68,-28.528959) (-67,-31.005313)
    (-66,-35.812536) (-65,-40.000000) (-64,-36.154945) (-63,-29.157220) (-62,-25.079216) (-61,-22.219328)
    (-60,-20.080788) (-59,-18.456160) (-58,-17.246142) (-57,-16.403870) (-56,-15.914516) (-55,-15.788819)
    (-54,-16.065181) (-53,-16.821914) (-52,-18.210601) (-51,-20.552578) (-50,-24.710268) (-49,-35.124287)
    (-48,-31.815954) (-47,-23.273813) (-46,-19.316588) (-45,-16.997731) (-44,-15.646877) (-43,-15.027002)
    (-42,-15.059619) (-41,-15.759635) (-40,-17.237795) (-39,-19.773725) (-38,-24.120152) (-37,-33.665198)
    (-36,-35.408931) (-35,-26.694083) (-34,-23.945477) (-33,-23.544979) (-32,-25.172897) (-31,-30.026622)
    (-30,-40.000000) (-29,-29.532635) (-28,-24.224073) (-27,-22.240384) (-26,-22.522872) (-25,-25.933426)
    (-24,-40.000000) (-23,-24.320128) (-22,-17.271863) (-21,-13.410099) (-20,-11.113261) (-19,-9.950616)
    (-18,-9.840278) (-17,-10.957341) (-16,-13.989765) (-15,-22.319060) (-14,-22.516297) (-13,-12.495983)
    (-12,-8.188824) (-11,-5.856844) (-10,-4.777832) (-9,-4.782659) (-8,-6.000739) (-7,-9.055066)
    (-6,-16.912908) (-5,-19.337314) (-4,-8.584887) (-3,-4.152690) (-2,-1.702008) (-1,-0.408084)
    (0,0.000000) (1,-0.408084) (2,-1.702008) (3,-4.152690) (4,-8.584887) (5,-19.337314)
    (6,-16.912908) (7,-9.055066) (8,-6.000739) (9,-4.782659) (10,-4.777832) (11,-5.856844)
    (12,-8.188824) (13,-12.495983) (14,-22.516297) (15,-22.319060) (16,-13.989765) (17,-10.957341)
    (18,-9.840278) (19,-9.950616) (20,-11.113261) (21,-13.410099) (22,-17.271863) (23,-24.320128)
    (24,-40.000000) (25,-25.933426) (26,-22.522872) (27,-22.240384) (28,-24.224073) (29,-29.532635)
    (30,-40.000000) (31,-30.026622) (32,-25.172897) (33,-23.544979) (34,-23.945477) (35,-26.694083)
    (36,-35.408931) (37,-33.665198) (38,-24.120152) (39,-19.773725) (40,-17.237795) (41,-15.759635)
    (42,-15.059619) (43,-15.027002) (44,-15.646877) (45,-16.997731) (46,-19.316588) (47,-23.273813)
    (48,-31.815954) (49,-35.124287) (50,-24.710268) (51,-20.552578) (52,-18.210601) (53,-16.821914)
    (54,-16.065181) (55,-15.788819) (56,-15.914516) (57,-16.403870) (58,-17.246142) (59,-18.456160)
    (60,-20.080788) (61,-22.219328) (62,-25.079216) (63,-29.157220) (64,-36.154945) (65,-40.000000)
    (66,-35.812536) (67,-31.005313) (68,-28.528959) (69,-27.096345) (70,-26.282678) (71,-25.894252)
    (72,-25.826782) (73,-26.018435) (74,-26.430577) (75,-27.038792) (76,-27.828365) (77,-28.792014)
    (78,-29.928908) (79,-31.244593) (80,-32.751751) (81,-34.471925) (82,-36.438690) (83,-38.703237)
    (84,-40.000000) (85,-40.000000) (86,-40.000000) (87,-40.000000) (88,-40.000000) (89,-40.000000)
    (90,-40.000000)
};
\addlegendentry{U-shape}

\addplot[solid, cyan, line width=0.75pt] coordinates {
    (-90,-40.000000) (-89,-40.000000) (-88,-40.000000) (-87,-40.000000) (-86,-40.000000) (-85,-40.000000)
    (-84,-40.000000) (-83,-38.708974) (-82,-36.448498) (-81,-34.487685) (-80,-32.775886) (-79,-31.280166)
    (-78,-29.979754) (-77,-28.862916) (-76,-27.925294) (-75,-27.169261) (-74,-26.604179) (-73,-26.247698)
    (-72,-26.128623) (-71,-26.292607) (-70,-26.813761) (-69,-27.820510) (-68,-29.562775) (-67,-32.637415)
    (-66,-39.265236) (-65,-40.000000) (-64,-33.573872) (-63,-28.068559) (-62,-24.602586) (-61,-22.147934)
    (-60,-20.355275) (-59,-19.076030) (-58,-18.248457) (-57,-17.862614) (-56,-17.953761) (-55,-18.616665)
    (-54,-20.057583) (-53,-22.770204) (-52,-28.420985) (-51,-40.000000) (-50,-26.102629) (-49,-20.268170)
    (-48,-16.944751) (-47,-14.846712) (-46,-13.586390) (-45,-13.032242) (-44,-13.181858) (-43,-14.156059)
    (-42,-16.301032) (-41,-20.685605) (-40,-34.949234) (-39,-23.809744) (-38,-16.648985) (-37,-13.104589)
    (-36,-11.118931) (-35,-10.184396) (-34,-10.172907) (-33,-11.172078) (-32,-13.580896) (-31,-18.841880)
    (-30,-40.000000) (-29,-18.345242) (-28,-12.608295) (-27,-9.768479) (-26,-8.416847) (-25,-8.209444)
    (-24,-9.168514) (-23,-11.708207) (-22,-17.527369) (-21,-35.187276) (-20,-15.307197) (-19,-10.222506)
    (-18,-7.816077) (-17,-6.925510) (-16,-7.317691) (-15,-9.202800) (-14,-13.644922) (-13,-29.714444)
    (-12,-16.266148) (-11,-9.824702) (-10,-7.007707) (-9,-5.992873) (-8,-6.476641) (-7,-8.826686)
    (-6,-15.200284) (-5,-23.209863) (-4,-9.451582) (-3,-4.510807) (-2,-1.839526) (-1,-0.440105)
    (0,0.000000) (1,-0.440105) (2,-1.839526) (3,-4.510807) (4,-9.451582) (5,-23.209863)
    (6,-15.200284) (7,-8.826686) (8,-6.476641) (9,-5.992873) (10,-7.007707) (11,-9.824702)
    (12,-16.266148) (13,-29.714444) (14,-13.644922) (15,-9.202800) (16,-7.317691) (17,-6.925510)
    (18,-7.816077) (19,-10.222506) (20,-15.307197) (21,-35.187276) (22,-17.527369) (23,-11.708207)
    (24,-9.168514) (25,-8.209444) (26,-8.416847) (27,-9.768479) (28,-12.608295) (29,-18.345242)
    (30,-40.000000) (31,-18.841880) (32,-13.580896) (33,-11.172078) (34,-10.172907) (35,-10.184396)
    (36,-11.118931) (37,-13.104589) (38,-16.648985) (39,-23.809744) (40,-34.949234) (41,-20.685605)
    (42,-16.301032) (43,-14.156059) (44,-13.181858) (45,-13.032242) (46,-13.586390) (47,-14.846712)
    (48,-16.944751) (49,-20.268170) (50,-26.102629) (51,-40.000000) (52,-28.420985) (53,-22.770204)
    (54,-20.057583) (55,-18.616665) (56,-17.953761) (57,-17.862614) (58,-18.248457) (59,-19.076030)
    (60,-20.355275) (61,-22.147934) (62,-24.602586) (63,-28.068559) (64,-33.573872) (65,-40.000000)
    (66,-39.265236) (67,-32.637415) (68,-29.562775) (69,-27.820510) (70,-26.813761) (71,-26.292607)
    (72,-26.128623) (73,-26.247698) (74,-26.604179) (75,-27.169261) (76,-27.925294) (77,-28.862916)
    (78,-29.979754) (79,-31.280166) (80,-32.775886) (81,-34.487685) (82,-36.448498) (83,-38.708974)
    (84,-40.000000) (85,-40.000000) (86,-40.000000) (87,-40.000000) (88,-40.000000) (89,-40.000000)
    (90,-40.000000)
};
\addlegendentry{Ring}

\addplot[solid, magenta, line width=0.75pt] coordinates {
    (-90,-40.000000) (-89,-40.000000) (-88,-40.000000) (-87,-40.000000) (-86,-40.000000) (-85,-40.000000)
    (-84,-40.000000) (-83,-40.000000) (-82,-40.000000) (-81,-40.000000) (-80,-40.000000) (-79,-39.308488)
    (-78,-37.955265) (-77,-36.768475) (-76,-35.739089) (-75,-34.863245) (-74,-34.141700) (-73,-33.579766)
    (-72,-33.187742) (-71,-32.982002) (-70,-32.987068) (-69,-33.239433) (-68,-33.794824) (-67,-34.743074)
    (-66,-36.242382) (-65,-38.614050) (-64,-40.000000) (-63,-40.000000) (-62,-40.000000) (-61,-40.000000)
    (-60,-36.918875) (-59,-34.272319) (-58,-32.541731) (-57,-31.479756) (-56,-30.998113) (-55,-31.098178)
    (-54,-31.867750) (-53,-33.538355) (-52,-36.719000) (-51,-40.000000) (-50,-40.000000) (-49,-37.941997)
    (-48,-33.104765) (-47,-30.382111) (-46,-28.837197) (-45,-28.192911) (-44,-28.412152) (-43,-29.652383)
    (-42,-32.443766) (-41,-38.870664) (-40,-40.000000) (-39,-34.315040) (-38,-29.343466) (-37,-26.720850)
    (-36,-25.438729) (-35,-25.250416) (-34,-26.234511) (-33,-28.901207) (-32,-35.442746) (-31,-40.000000)
    (-30,-30.244857) (-29,-25.475567) (-28,-23.073232) (-27,-22.107438) (-26,-22.415736) (-25,-24.282015)
    (-24,-29.033935) (-23,-40.000000) (-22,-28.615539) (-21,-22.639593) (-20,-19.761964) (-19,-18.545721)
    (-18,-18.735052) (-17,-20.634829) (-16,-25.873031) (-15,-40.000000) (-14,-22.883942) (-13,-17.388307)
    (-12,-14.648201) (-11,-13.523426) (-10,-13.881350) (-9,-16.291743) (-8,-24.180428) (-7,-23.242692)
    (-6,-12.758348) (-5,-7.761282) (-4,-4.587979) (-3,-2.452235) (-2,-1.055215) (-1,-0.259089)
    (0,0.000000) (1,-0.259089) (2,-1.055215) (3,-2.452235) (4,-4.587979) (5,-7.761282)
    (6,-12.758348) (7,-23.242692) (8,-24.180428) (9,-16.291743) (10,-13.881350) (11,-13.523426)
    (12,-14.648201) (13,-17.388307) (14,-22.883942) (15,-40.000000) (16,-25.873031) (17,-20.634829)
    (18,-18.735052) (19,-18.545721) (20,-19.761964) (21,-22.639593) (22,-28.615539) (23,-40.000000)
    (24,-29.033935) (25,-24.282015) (26,-22.415736) (27,-22.107438) (28,-23.073232) (29,-25.475567)
    (30,-30.244857) (31,-40.000000) (32,-35.442746) (33,-28.901207) (34,-26.234511) (35,-25.250416)
    (36,-25.438729) (37,-26.720850) (38,-29.343466) (39,-34.315040) (40,-40.000000) (41,-38.870664)
    (42,-32.443766) (43,-29.652383) (44,-28.412152) (45,-28.192911) (46,-28.837197) (47,-30.382111)
    (48,-33.104765) (49,-37.941997) (50,-40.000000) (51,-40.000000) (52,-36.719000) (53,-33.538355)
    (54,-31.867750) (55,-31.098178) (56,-30.998113) (57,-31.479756) (58,-32.541731) (59,-34.272319)
    (60,-36.918875) (61,-40.000000) (62,-40.000000) (63,-40.000000) (64,-40.000000) (65,-38.614050)
    (66,-36.242382) (67,-34.743074) (68,-33.794824) (69,-33.239433) (70,-32.987068) (71,-32.982002)
    (72,-33.187742) (73,-33.579766) (74,-34.141700) (75,-34.863245) (76,-35.739089) (77,-36.768475)
    (78,-37.955265) (79,-39.308488) (80,-40.000000) (81,-40.000000) (82,-40.000000) (83,-40.000000)
    (84,-40.000000) (85,-40.000000) (86,-40.000000) (87,-40.000000) (88,-40.000000) (89,-40.000000)
    (90,-40.000000)
};
\addlegendentry{Diagonal stripe}

\addplot[solid, blue!60!black, line width=0.75pt] coordinates {
    (-90,-16.511518) (-89,-16.511515) (-88,-16.511466) (-87,-16.511256) (-86,-16.510692) (-85,-16.509507)
    (-84,-16.507369) (-83,-16.503894) (-82,-16.498672) (-81,-16.491306) (-80,-16.481466) (-79,-16.468969)
    (-78,-16.453867) (-77,-16.436556) (-76,-16.417886) (-75,-16.399269) (-74,-16.382758) (-73,-16.371092)
    (-72,-16.367687) (-71,-16.376560) (-70,-16.402179) (-69,-16.449245) (-68,-16.522392) (-67,-16.625826)
    (-66,-16.762910) (-65,-16.935709) (-64,-17.144560) (-63,-17.387742) (-62,-17.661400) (-61,-17.959942)
    (-60,-18.277103) (-59,-18.607766) (-58,-18.950299) (-57,-19.308755) (-56,-19.693984) (-55,-20.122812)
    (-54,-20.614817) (-53,-21.186531) (-52,-21.842735) (-51,-22.564569) (-50,-23.296997) (-49,-23.947730)
    (-48,-24.421161) (-47,-24.690793) (-46,-24.850514) (-45,-25.088433) (-44,-25.630482) (-43,-26.730908)
    (-42,-28.729574) (-41,-32.054968) (-40,-34.796500) (-39,-30.721014) (-38,-26.126561) (-37,-22.887687)
    (-36,-20.606135) (-35,-18.998014) (-34,-17.896981) (-33,-17.191160) (-32,-16.788006) (-31,-16.594938)
    (-30,-16.511518) (-29,-16.434588) (-28,-16.273853) (-27,-15.964251) (-26,-15.462965) (-25,-14.743129)
    (-24,-13.806769) (-23,-12.707418) (-22,-11.546800) (-21,-10.441279) (-20,-9.489511) (-19,-8.760264)
    (-18,-8.294777) (-17,-8.112516) (-16,-8.213930) (-15,-8.577620) (-14,-9.150938) (-13,-9.834877)
    (-12,-10.470008) (-11,-10.841216) (-10,-10.727011) (-9,-10.001200) (-8,-8.724832) (-7,-7.118479)
    (-6,-5.433021) (-5,-3.855797) (-4,-2.497186) (-3,-1.413397) (-2,-0.630088) (-1,-0.157732)
    (0,0.000000) (1,-0.157732) (2,-0.630088) (3,-1.413397) (4,-2.497186) (5,-3.855797)
    (6,-5.433021) (7,-7.118479) (8,-8.724832) (9,-10.001200) (10,-10.727011) (11,-10.841216)
    (12,-10.470008) (13,-9.834877) (14,-9.150938) (15,-8.577620) (16,-8.213930) (17,-8.112516)
    (18,-8.294777) (19,-8.760264) (20,-9.489511) (21,-10.441279) (22,-11.546800) (23,-12.707418)
    (24,-13.806769) (25,-14.743129) (26,-15.462965) (27,-15.964251) (28,-16.273853) (29,-16.434588)
    (30,-16.511518) (31,-16.594938) (32,-16.788006) (33,-17.191160) (34,-17.896981) (35,-18.998014)
    (36,-20.606135) (37,-22.887687) (38,-26.126561) (39,-30.721014) (40,-34.796500) (41,-32.054968)
    (42,-28.729574) (43,-26.730908) (44,-25.630482) (45,-25.088433) (46,-24.850514) (47,-24.690793)
    (48,-24.421161) (49,-23.947730) (50,-23.296997) (51,-22.564569) (52,-21.842735) (53,-21.186531)
    (54,-20.614817) (55,-20.122812) (56,-19.693984) (57,-19.308755) (58,-18.950299) (59,-18.607766)
    (60,-18.277103) (61,-17.959942) (62,-17.661400) (63,-17.387742) (64,-17.144560) (65,-16.935709)
    (66,-16.762910) (67,-16.625826) (68,-16.522392) (69,-16.449245) (70,-16.402179) (71,-16.376560)
    (72,-16.367687) (73,-16.371092) (74,-16.382758) (75,-16.399269) (76,-16.417886) (77,-16.436556)
    (78,-16.453867) (79,-16.468969) (80,-16.481466) (81,-16.491306) (82,-16.498672) (83,-16.503894)
    (84,-16.507369) (85,-16.509507) (86,-16.510692) (87,-16.511256) (88,-16.511466) (89,-16.511515)
    (90,-16.511518)
};
\addlegendentry{Plus shape}

\addplot[solid, orange!80!black, line width=0.75pt] coordinates {
    (-90,-40.000000) (-89,-40.000000) (-88,-40.000000) (-87,-40.000000) (-86,-40.000000) (-85,-40.000000)
    (-84,-40.000000) (-83,-38.636542) (-82,-36.324789) (-81,-34.289146) (-80,-32.472388) (-79,-30.833920)
    (-78,-29.343994) (-77,-27.980246) (-76,-26.725523) (-75,-25.566464) (-74,-24.492533) (-73,-23.495359)
    (-72,-22.568256) (-71,-21.705882) (-70,-20.903981) (-69,-20.159201) (-68,-19.468949) (-67,-18.831284)
    (-66,-18.244846) (-65,-17.708795) (-64,-17.222775) (-63,-16.786898) (-62,-16.401735) (-61,-16.068333)
    (-60,-15.788240) (-59,-15.563558) (-58,-15.397016) (-57,-15.292080) (-56,-15.253105) (-55,-15.285550)
    (-54,-15.396289) (-53,-15.594044) (-52,-15.890052) (-51,-16.299054) (-50,-16.840880) (-49,-17.543059)
    (-48,-18.445408) (-47,-19.608685) (-46,-21.132630) (-45,-23.199022) (-44,-26.198102) (-43,-31.264103)
    (-42,-40.000000) (-41,-34.245139) (-40,-27.144949) (-39,-23.248778) (-38,-20.572892) (-37,-18.563986)
    (-36,-16.989114) (-35,-15.729477) (-34,-14.717704) (-33,-13.913436) (-32,-13.292327) (-31,-12.840662)
    (-30,-12.552725) (-29,-12.429766) (-28,-12.480161) (-27,-12.720835) (-26,-13.180503) (-25,-13.906225)
    (-24,-14.977024) (-23,-16.535160) (-22,-18.871133) (-21,-22.732826) (-20,-31.491601) (-19,-32.075602)
    (-18,-21.814041) (-17,-16.980490) (-16,-13.743106) (-15,-11.300624) (-14,-9.345837) (-13,-7.727905)
    (-12,-6.361519) (-11,-5.193774) (-10,-4.189668) (-9,-3.324935) (-8,-2.582156) (-7,-1.948512)
    (-6,-1.414407) (-5,-0.972592) (-4,-0.617595) (-3,-0.345329) (-2,-0.152836) (-1,-0.038114)
    (0,0.000000) (1,-0.038114) (2,-0.152836) (3,-0.345329) (4,-0.617595) (5,-0.972592)
    (6,-1.414407) (7,-1.948512) (8,-2.582156) (9,-3.324935) (10,-4.189668) (11,-5.193774)
    (12,-6.361519) (13,-7.727905) (14,-9.345837) (15,-11.300624) (16,-13.743106) (17,-16.980490)
    (18,-21.814041) (19,-32.075602) (20,-31.491601) (21,-22.732826) (22,-18.871133) (23,-16.535160)
    (24,-14.977024) (25,-13.906225) (26,-13.180503) (27,-12.720835) (28,-12.480161) (29,-12.429766)
    (30,-12.552725) (31,-12.840662) (32,-13.292327) (33,-13.913436) (34,-14.717704) (35,-15.729477)
    (36,-16.989114) (37,-18.563986) (38,-20.572892) (39,-23.248778) (40,-27.144949) (41,-34.245139)
    (42,-40.000000) (43,-31.264103) (44,-26.198102) (45,-23.199022) (46,-21.132630) (47,-19.608685)
    (48,-18.445408) (49,-17.543059) (50,-16.840880) (51,-16.299054) (52,-15.890052) (53,-15.594044)
    (54,-15.396289) (55,-15.285550) (56,-15.253105) (57,-15.292080) (58,-15.397016) (59,-15.563558)
    (60,-15.788240) (61,-16.068333) (62,-16.401735) (63,-16.786898) (64,-17.222775) (65,-17.708795)
    (66,-18.244846) (67,-18.831284) (68,-19.468949) (69,-20.159201) (70,-20.903981) (71,-21.705882)
    (72,-22.568256) (73,-23.495359) (74,-24.492533) (75,-25.566464) (76,-26.725523) (77,-27.980246)
    (78,-29.343994) (79,-30.833920) (80,-32.472388) (81,-34.289146) (82,-36.324789) (83,-38.636542)
    (84,-40.000000) (85,-40.000000) (86,-40.000000) (87,-40.000000) (88,-40.000000) (89,-40.000000)
    (90,-40.000000)
};
\addlegendentry{Corner block}

\addplot[solid, yellow!70!black, line width=0.75pt] coordinates {
    (-90,-40.000000) (-89,-40.000000) (-88,-40.000000) (-87,-40.000000) (-86,-40.000000) (-85,-40.000000)
    (-84,-40.000000) (-83,-38.680260) (-82,-36.399385) (-81,-34.408695) (-80,-32.654775) (-79,-31.101364)
    (-78,-29.723636) (-77,-28.504818) (-76,-27.434135) (-75,-26.505561) (-74,-25.717088) (-73,-25.070407)
    (-72,-24.570923) (-71,-24.228145) (-70,-24.056563) (-69,-24.077253) (-68,-24.320732) (-67,-24.832183)
    (-66,-25.681615) (-65,-26.985696) (-64,-28.962030) (-63,-32.099294) (-62,-37.983071) (-61,-40.000000)
    (-60,-36.897303) (-59,-31.125547) (-58,-27.890708) (-57,-25.832792) (-56,-24.538263) (-55,-23.849069)
    (-54,-23.718861) (-53,-24.178802) (-52,-25.352592) (-51,-27.544681) (-50,-31.607562) (-49,-40.000000)
    (-48,-38.673220) (-47,-30.188565) (-46,-26.350448) (-45,-24.228981) (-44,-23.183832) (-43,-23.033754)
    (-42,-23.812881) (-41,-25.800025) (-40,-29.920766) (-39,-40.000000) (-38,-35.062361) (-37,-27.447783)
    (-36,-24.004436) (-35,-22.308952) (-34,-21.838917) (-33,-22.546510) (-32,-24.772440) (-31,-29.931644)
    (-30,-40.000000) (-29,-29.435684) (-28,-23.805641) (-27,-21.165009) (-26,-20.146116) (-25,-20.498403)
    (-24,-22.497241) (-23,-27.577497) (-22,-40.000000) (-21,-26.458427) (-20,-20.912533) (-19,-18.352829)
    (-18,-17.493958) (-17,-18.178329) (-16,-20.965764) (-15,-29.182123) (-14,-29.377000) (-13,-19.465988)
    (-12,-15.404311) (-11,-13.516407) (-10,-13.227569) (-9,-14.774075) (-8,-19.956980) (-7,-31.761029)
    (-6,-14.509275) (-5,-8.570309) (-4,-5.004357) (-3,-2.657550) (-2,-1.139424) (-1,-0.279238)
    (0,0.000000) (1,-0.279238) (2,-1.139424) (3,-2.657550) (4,-5.004357) (5,-8.570309)
    (6,-14.509275) (7,-31.761029) (8,-19.956980) (9,-14.774075) (10,-13.227569) (11,-13.516407)
    (12,-15.404311) (13,-19.465988) (14,-29.377000) (15,-29.182123) (16,-20.965764) (17,-18.178329)
    (18,-17.493958) (19,-18.352829) (20,-20.912533) (21,-26.458427) (22,-40.000000) (23,-27.577497)
    (24,-22.497241) (25,-20.498403) (26,-20.146116) (27,-21.165009) (28,-23.805641) (29,-29.435684)
    (30,-40.000000) (31,-29.931644) (32,-24.772440) (33,-22.546510) (34,-21.838917) (35,-22.308952)
    (36,-24.004436) (37,-27.447783) (38,-35.062361) (39,-40.000000) (40,-29.920766) (41,-25.800025)
    (42,-23.812881) (43,-23.033754) (44,-23.183832) (45,-24.228981) (46,-26.350448) (47,-30.188565)
    (48,-38.673220) (49,-40.000000) (50,-31.607562) (51,-27.544681) (52,-25.352592) (53,-24.178802)
    (54,-23.718861) (55,-23.849069) (56,-24.538263) (57,-25.832792) (58,-27.890708) (59,-31.125547)
    (60,-36.897303) (61,-40.000000) (62,-37.983071) (63,-32.099294) (64,-28.962030) (65,-26.985696)
    (66,-25.681615) (67,-24.832183) (68,-24.320732) (69,-24.077253) (70,-24.056563) (71,-24.228145)
    (72,-24.570923) (73,-25.070407) (74,-25.717088) (75,-26.505561) (76,-27.434135) (77,-28.504818)
    (78,-29.723636) (79,-31.101364) (80,-32.654775) (81,-34.408695) (82,-36.399385) (83,-38.680260)
    (84,-40.000000) (85,-40.000000) (86,-40.000000) (87,-40.000000) (88,-40.000000) (89,-40.000000)
    (90,-40.000000)
};
\addlegendentry{Checkerboard}

\end{axis}
\end{tikzpicture}

%% file: sum_rate_vs_snr.tex
\begin{tikzpicture}
\begin{axis}[
    width=1\columnwidth,
    height=0.37\textwidth,
    xlabel={SNR [$\rm{dB}$]},
    ylabel={Sum-rate [$\rm{bits/Hz}$]},
    grid=major,
    legend style={
        at={(0.02,0.98)},
        anchor=north west,
        font=\scriptsize,
        inner sep=0.5mm,
        legend cell align={left},
        legend columns=1,
        /tikz/column 1/.style={column sep=-1pt,},
        /tikz/column 2/.style={column sep=0pt,},
        /tikz/row 1/.style={row sep=-2pt,},
        /tikz/row 2/.style={row sep=-2pt,},
        /tikz/row 3/.style={row sep=-2pt,},
        /tikz/row 4/.style={row sep=-2pt,},
        /tikz/row 5/.style={row sep=-2pt,},
        /tikz/row 6/.style={row sep=-2pt,},
    },
    tick label style={font=\footnotesize},
    xlabel style={font=\footnotesize},
    ylabel style={font=\footnotesize},
    title style={font=\footnotesize},
    xmin=-10, xmax=30,
    xtick={-10,-5,0,5,10,15,20,25,30},
    ymin=0, ymax=90,
    ytick={0,10,20,30,40,50,60,70,80,90},
    legend entries={
        Shape-adaptive RHS,
        Fixed RHS (Full aperture),
        Fixed RHS (Centered rectangle),
        Random-shape RHS,
        RIS-like (phase-only),
        No surface
    }
]

\addplot[mark=o, solid, blue, line width=0.7pt] coordinates {
    (-10,10.173696086181401) (-5,16.529311964480058) (0,24.041873055641155)
    (5,32.625758144004074) (10,42.125220956871907) (15,51.926722100844245)
    (20,61.818022906829938) (25,72.006287156479232) (30,81.863413100306232)
};

\addplot[mark=square*, solid, orange, line width=0.7pt] coordinates {
    (-10,4.600472734689844) (-5,9.567823252002565) (0,16.820170992840314)
    (5,25.575100626131370) (10,35.179676239103493) (15,45.035620053956023)
    (20,54.904733328847321) (25,65.027944546848389) (30,74.675264533781260)
};

\addplot[mark=diamond*, solid, yellow!80!black, line width=0.7pt] coordinates {
    (-10,4.662366182557888) (-5,9.106590570650978) (0,15.460266758855539)
    (5,23.437921199373580) (10,32.497998741907480) (15,42.507479711337730)
    (20,52.280557774008955) (25,62.333668809441868) (30,72.132235890592824)
};

\addplot[mark=none, dashed, purple, line width=0.7pt] coordinates {
    (-10,5.748506235994840) (-5,10.416218567814969) (0,17.442536706755664)
    (5,25.310456249885732) (10,35.325172038212749) (15,44.172334925244321)
    (20,54.094996412541065) (25,64.304573157396334) (30,74.749366945254451)
};

\addplot[mark=triangle*, solid, green!70!black, line width=0.7pt] coordinates {
    (-10,4.600472734689844) (-5,9.567823252002565) (0,16.820170992840314)
    (5,25.575100626131370) (10,35.179676239103493) (15,45.035620053956023)
    (20,54.904733328847321) (25,65.027944546848389) (30,74.675264533781260)
};

\addplot[mark=x, solid, cyan, line width=0.7pt] coordinates {
    (-10,0.946385846392669) (-5,2.614033628613834) (0,5.721785620152873)
    (5,10.158418595661992) (10,16.514069298826360) (15,23.094445730483066)
    (20,30.491482266286145) (25,37.313320203062325) (30,45.261716652832355)
};

\end{axis}
\end{tikzpicture}